# FOCUSING PROPERTIES OF MUSHROOM MICROLENSES

A.V. Boriskin[1], S.V. Boriskina[2], T. Benson[2], P.Sewell[2], A.I. Nosich[1,2]

[1] *Institute of Radiophysics and Electronics NASU, ul. Proskury 12, Kharkov 61085, Ukraine*
e-mail: a_boriskin@yahoo.com, web-page: www.ire.kharkov.ua/dep12/MOCA
[2] *G. Green Institute for Electromagnetic Research, University of Nottingham, NG7 2RD, UK*

**Abstract** - Focusing properties of a novel type photoresist microlens are studied. A specific character of the microlens is its mushroom shape. Recently it was predicted and experimentally revealed that such a lens integrated with a light-emitting diode is capable of enhancing its output efficiency and directivity [1]. In our paper we analyze the electromagnetic performance of the mushroom lens by applying a mathematically rigorous method of boundary integral equations. Numerical results are presented for the mushroom lens illuminated with a plane E-polarized wave and include figures describing the evolution of the lens focal spot and near field maps.

Microlenses of different sizes and shapes are used to enhance the brightness and output pattern of light-emitting diodes and to improve the sensitivity of light detectors [1-3]. One of the novel and promising types of such microlenses is a lens of mushroom shape (Fig. 1). As it was recently shown by Heremans [1], the microlenses with a dome located at some distance from the substrate (so-called mushroom microlenses) provide better collimation of output light comparing with commonly used hemispherical ones. To explain this, the mushroom lens was considered as a set of 3 planoconvex lenses (Fig. 2) whose focusing properties are estimated in the ray-tracing approximation valid only until the wavelength is much smaller than the lens diameter. In such a way, though a mushroom lenses fabrication technique based on reflow is proven to be effective [1, 3], the electromagnetic behavior of such lenses is still far from clear.

To study the electromagnetic performance of such a lens we use the Muller boundary integral equation method, which is known as mathematically rigorous and accurate tool of electromagnetic analysis of dielectric scatterers. The specific shape of the lens profile is described by cubic splines.

The research is aimed at the study the focusing properties of mushroom microlenses and check of the equivalence validity between the mushroom lens and the lensset proposed by Heremans [1].

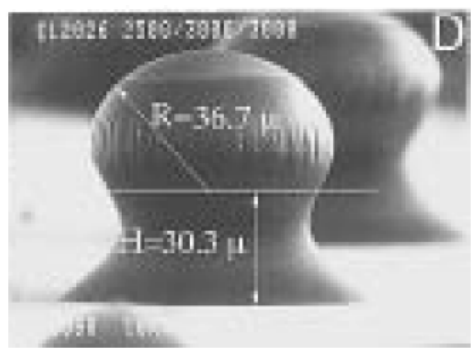

Fig. 1. A photograph of a photoresist mushroom microlens borrowed from [1].

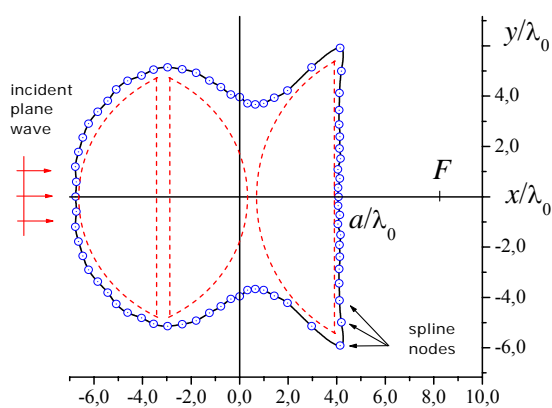

Fig. 2. Profile of the mushroom lens shown in Fig. 1 interpolated with cubic splines. With dotted lines the three planoconvex lenses proposed in [1] as an equivalent lensset are indicated.

## OUTLINE OF SOLUTION

We analyze the lens electromagnetic performance by applying the method of Muller's boundary integral equations (MBIEs). When combined with the trigonometric Galerkin discretization scheme, it enables one to obtain a set of coupled matrix equations of the Fredholm second kind where unknowns are the expansion coefficients of the layer potentials over the lens contour. Numerical algorithm based on the MBIEs possesses high accuracy and fast convergence [4].

To represent the mushroom lens profile we use cubic splines, which guarantee flexibility of the contour while providing the continuous first and second derivatives necessary for the stability of the algorithm based on the MBIEs. The values of the spline nodes coordinates were set to describe the lens given in Fig. 1. As we parameterize the contour with the aid of the polar angle, we are limited to the so-called star-like shapes: the ones for whom a straight line from the origin crosses the contour only once. The possible shapes of the mushroom lenses are determined by the relation between the surface tension of the photoresist during reflow and the internal pressure given by Eq. 2 in [1]. Fortunately, most of them can be characterized in such a way. Nevertheless, one has to be careful when placing an origin inside the lens. Besides, a number of spline nodes, $N_p$, has to be chosen quite large to describe the specific profile of the mushroom lens. We assume the angular mesh to be uniform. Non-uniform angular mesh can be used to decrease the number of spline nodes, though it does not fasten the algorithm.

## NUMERICAL RESULTS

We consider the photoresist (n ≈ 1.6) mushroom lens illuminated with a plane $E$-polarized wave in symmetrical manner. Geometrical optics (GO) says that a lensset of three lenses shown in Fig. 2 collects parallel rays in a focal spot located at point $F \approx (2a/\lambda_0, 0)$. It is well known that any finite lens has a focal spot of a finite size. Its shape and location depend on the lens material and profile as well as on the polarization and the angle of incidence of the plane wave [5]. To study the focusing properties of the mushroom lens, we plot the x-coordinate of the focal domain, namely the point with the maximum field intensity $I = |E_z(x,0)|^2$, vs. the normalized frequency parameter (Fig. 3).

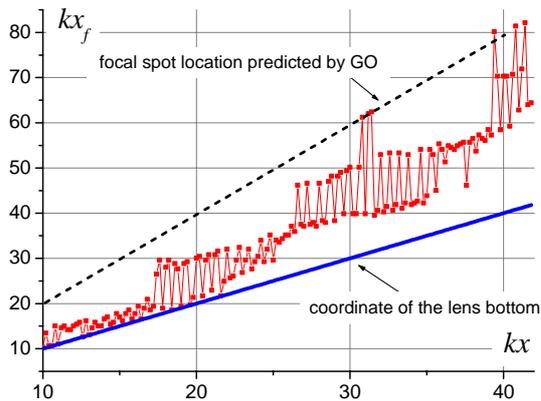
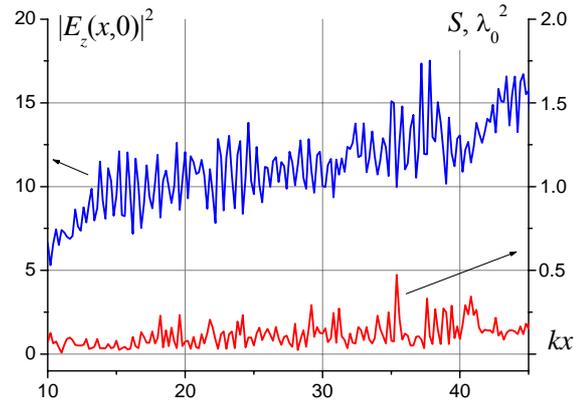

Fig. 3. The *x*-coordinate of the point with the maximum field intensity vs. the normalized frequency parameter.

Fig. 4. The field intensity in the focal spot (left axis) and the focal domain size (right axis) vs. the normalized frequency parameter.

One can see that the focal spot is not located where it is predicted in GO and migrates along the lens axis in some specific manner: for the lens of about 2 wavelengths in free space it is located close to the lens bottom while for larger ones it shifts back and forward along the lens axis. The focal domain size, S, in $\lambda_0^2$, and maximum field intensity also change quite rapidly with frequency growth (Fig. 4). The multiple ripples in Fig. 3 and 4 show the dominant role of internal resonances in the mushroom lens behavior. Fig. 5 presents the near field maps for the mushroom lens and a hemispherical one with the same bottom size. As one can see, in general the mushroom lens behaves as it was predicted by using the lensset representation: its focal domain is located closer to the lens bottom than for the classical hemielliptical lens. Though, its focal domain has an elongated shape. As it is seen from Fig. 3 and 4, the spot size and location depend on lens size in terms of wavelength and thus can be controlled by the change of lens parameters. As to the field intensity in the focal spots of two lenses, it is almost the same.

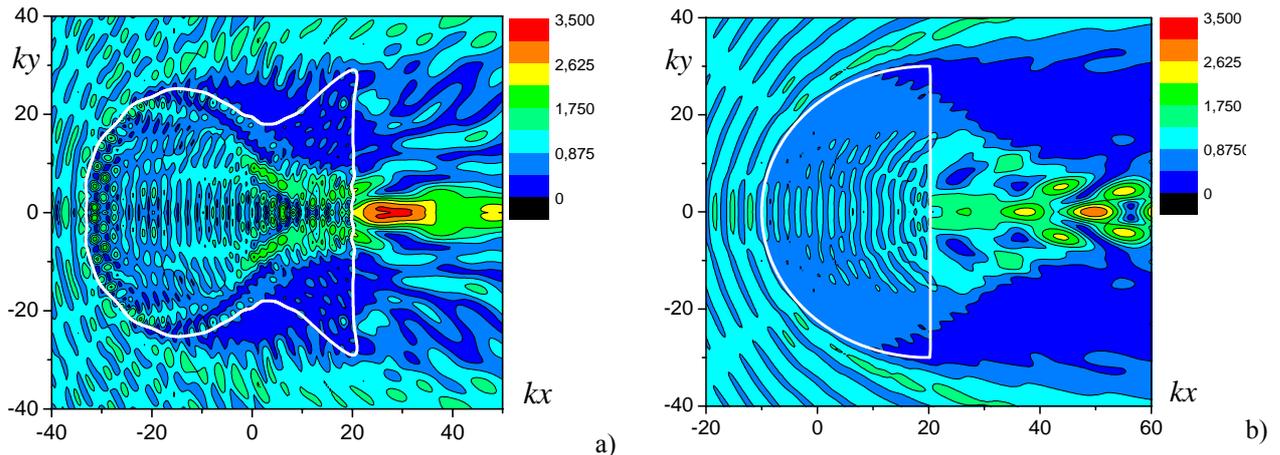

Fig. 5. Near field maps of the mushroom and hemielliptical lenses with equal bottoms illuminated with a plane *E*-polarized wave in symmetrical manner.

## CONCLUSIONS

The photoresist lens focusing properties have been studied in accurate manner by applying the Muller boundary integral equation method. The shift of the focal domain towards the lens bottom predicted in [1] by using the equivalent lensset of three planoconvex lenses has been observed. Besides, the important role of the internal resonances in the electromagnetic performance of finite size mushroom microlenses has been demonstrated.

## ACKNOWLEDGMENT

This research work has been supported by the Royal Society via the Grant IJP-2004/R1-FS.